\documentclass[
 prl, twocolumn,
letterpaper, superscriptaddress,
longbibliography]
{revtex4-1}
\usepackage{amsmath, amssymb}
\usepackage[dvipsnames]{xcolor}
\usepackage{hyperref, cleveref}
\usepackage{graphicx}

\newcommand{\T}[0]{\mathcal{T}}

\begin{document}
\title{Nodal lines in a honeycomb plasmonic crystal with synthetic spin}

\author{Sang Hyun Park}
\affiliation{Department of Electrical \& Computer Engineering, University of Minnesota, Minneapolis, Minnesota, 55455, USA}
\author{E. J. Mele}
\affiliation{Department of Physics and Astronomy, University of Pennsylvania, Philadelphia, Pennsylvania, 19104, USA}
\author{Tony Low}
\email{tlow@umn.edu}
\affiliation{Department of Electrical \& Computer Engineering, University of Minnesota, Minneapolis, Minnesota, 55455, USA}

\begin{abstract}
	We analyze a plasmonic model on a honeycomb lattice of metallic nanodisks that hosts nodal lines protected by local symmetries.  Using both continuum and tight-binding models, we show that a combination of a synthetic time-reversal symmetry, inversion symmetry, and particle-hole symmetry enforce the existence of nodal lines enclosing the $\mathrm{K}$ and $\mathrm{K}'$ points. The nodal lines are not directly gapped even when these symmetries are weakly broken. The existence of the nodal lines is verified using full-wave electromagnetic simulations. We also show that the degeneracies at nodal lines can be relieved by introducing a Kekul\'e distortion that acts to mix the nodal lines near the $\mathrm{K},\mathrm{K}'$ points. Our findings open pathways for designing novel plasmonic and photonic devices without reliance on complex symmetry engineering, presenting a convenient platform for studying nodal structures in two-dimensional systems.
\end{abstract}
\maketitle

Gapless phases of quantum matter have recently gained widespread interest\cite{Armitage2018}. Following the seminal observation of Dirac points in graphene\cite{CastroNeto2009}, the search for gapless phases has led to the discovery of a wide variety of momentum-space degeneracies such as Weyl points\cite{Wan2011, Xu2015}, nodal lines\cite{Burkov2011, Xu2011,Fang2015, Young2015, Schoop2016, Bian2016, Bian2016a, Feng2017a}, and nodal surfaces\cite{Wu2018,Qiu2024}. Among these, nodal lines are particularly intriguing because of their higher dimensionality than Dirac and Weyl points, allowing for various geometric configurations. Nodal rings\cite{Kim2015, Yu2015}, nodal chains\cite{Bzdusek2016, Yu2017}, nodal links\cite{Chang2017, Yan2017}, and nodal knots\cite{Ezawa2017, Bi2017} are just a few examples of the different configurations that have been reported. Beyond electronic systems, photonic and plasmonic platforms have also been used to explore gapless phases such as Dirac points\cite{Weick2013}, Weyl points\cite{Lu2013, Lin2016}, and nodal lines\cite{Xia2019, Gao2018c, Lin2017b} due to their tunable band structures and experimental accessibility. 

Nodal lines can be grouped into two distinct classes depending on whether they can be removed without or with lowering symmetries of a system. In the first class, line degeneracies are formed when bands in orthogonal subspaces of some symmetry operation have accidental crossings. These can be removed by continuously lifting the crossing without lowering the symmetry of the system. In the second class, the crossing is instead enforced by a symmetry. A minimum of four bands are entangled by their pinning to two-fold degeneracies on Brillouin zone (BZ) boundaries. Their intersections on lines in the interior of the BZ cannot be removed without lowering the symmetry. Nonsymmorphic symmetries play a central role in realizing this latter type of symmetry-enforced nodal line as they guarantee that the bands swap pairs between two time-reversal invariant momenta in the BZ. Most demonstrations of symmetry-enforced nodal lines in both electronic and photonic crystals in two and three dimensions rely on the existence of nonsymmorphic symmetries. 

Here we identify a honeycomb plasmonic crystal that hosts nodal lines without requiring nonsymmorphic symmetries. The plasmonic crystal is constructed from metallic nanodisks which each support localized multipolar plasmonic modes that play the role of orbitals \cite{Park2024, Li2023}. Coupling between the doubly degenerate hexapolar modes of the disk results in a four-band theory where each band can be classified by the eigenvalues $\lambda=\pm 1$ of a symmetry operation encoding the rotational symmetry of the disks. In the absence of additional interactions, there would be four-fold degenerate Dirac points at the $\mathrm{K}$ and $\mathrm{K}'$-points. An effective second-neighbor coupling mediated by an inter-orbital matrix element results in a splitting of the four-fold degenerate Dirac points into two-fold degenerate partner Dirac points while leaving behind a nodal line at zero energy enclosing the $\mathrm{K, K}'$ points. We find that a synthetic time-reversal symmetry($\mathcal{T}$), inversion symmetry($\mathcal{P}$), and particle-hole symmetry($\mathcal{C}$) of the system require pairs of bands to  degenerate at $\mathrm{K, K}'$ ($\Gamma,\mathrm{M}$) with identical(opposite) values of $\lambda$, thus protecting the nodal loop centered around $\mathrm{K}$, $\mathrm{K}'$. Using both tight-binding and full-wave electromagnetic simulations, we show that the nodal loop survives even in the presence of perturbations that weakly break the $\mathcal{T}$ and $\mathcal{P}$ symmetries of the system. Finally, we show that a gapped nodal line state can be realized by introducing a Kekul\'e distortion\cite{Hou2007}, mirroring the well-known chiral-symmetry-breaking mechanism for gapped Dirac points in Kekul\'e-distorted graphene\cite{Gutierrez2016,Gamayun2018,Bao2021}. As such, the proposed honeycomb plasmonic crystal provides a natural platform for investigating the properties of nodal lines in two-dimensional photonic systems without the need to engineer nonsymmorphic symmetries explicitly.

 \begin{figure}
 	\centering
 	\includegraphics{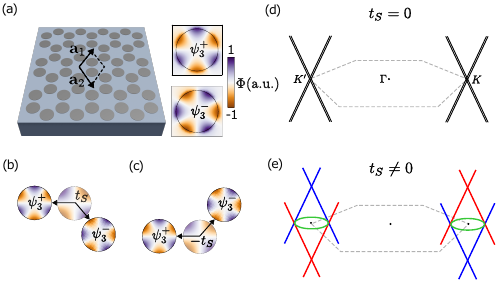}
	\caption{\textbf{Interactions in a honeycomb plasmonic crystal}. (a) Schematic representation of the honeycomb plasmonic crystal comprised of metallic nanodisks. The electric potential of the doubly degenerate hexapolar modes of the metallic nanodisks is shown to the right. (b,c) Effective second neighbor interactions between the hexapolar modes. The effective second-neighbor interaction mediated by an inter-orbital interaction involving the quadrupole modes is shown. (d) Dirac cones without the effective second neighbor interaction. (e) Nodal lines are formed when the effective second-neighbor interaction is introduced.}
 	\label{fig:1}
 \end{figure}
 
{\color{NavyBlue}\emph{Continuum model and symmetries}}.---The isolated metallic nanodisks in our model each support a ladder of multipolar modes among which the six lowest energy modes are doubly degenerate dipole, quadrupole, and hexapole modes. Here, we will be focusing on the plasmonic band structure formed by coupling between the hexapole modes ($\psi_3^+,\psi_3^-$) since they share the $C_3$ symmetry of the honeycomb lattice (see \cref{fig:1}a). The coupling coefficient between modes can be deduced by analyzing the overlap between their electric potentials\cite{Park2024,Li2023}. From the potentials shown in \cref{fig:1}a, we find that the coupling coefficient between $\psi_3^+$ modes is positive whereas for the $\psi_3^-$ modes, it is negative. There is no direct coupling between $\psi_3^+$ and $\psi_3^-$ modes. Therefore, when the metallic nanodisks are arranged into a honeycomb lattice, the hexapolar modes will couple to form two Dirac cones at each of the $\mathrm{K}$ and $\mathrm{K}'$ points. Assuming that the coupling coefficients for $\psi_3^\pm$ have opposite signs and equal magnitude, the two Dirac cones will be inverted versions of each other (see \cref{fig:1}c). The low-energy Hamiltonian around the $\mathrm{K}$, $\mathrm{K}'$ points is a four-band model that may be written as
\begin{equation}
	H_0(\mathbf{K}^\xi+\mathbf{q})=\hbar v\left(\xi \sigma_x q_x+\sigma_yq_y\right)s_z
\end{equation}
where $\xi=\pm 1$ denote the $\mathrm{K}/\mathrm{K}'$ valleys, $\boldsymbol{\sigma}$ acts on the sublattice degrees of freedom $\{|A\rangle,|B\rangle\}$, and $\mathbf{s}$ acts on the hexapole basis states $\{|\psi_3^+\rangle,|\psi_3^-\rangle\}$ which functions as an internal synthetic spin degree of freedom.

The plasmonic lattice has a synthetic time-reversal symmetry, inversion symmetry, and particle-hole symmetry. The synthetic time-reversal symmetry satisfies $\T H(\mathbf{k})\T^{-1}=H(-\mathbf{k})$ and is represented as $\T=i\sigma_z s_yK$ where $\T^2=-1$. Note the additional $\sigma_z$ term when compared to the conventional time-reversal operator that is required because the two synthetic spin states have coupling coefficients with opposite signs. Inversion symmetry satisfies $\mathcal{P}H(\mathbf{k})\mathcal{P}^{-1}=H(-\mathbf{k})$ and is represented as $\mathcal{P}=\sigma_x$. Finally, particle-hole symmetry satisfies $\mathcal{C}H(\mathbf{k})\mathcal{C}^{-1}=-H(-\mathbf{k})$ and is represented as $\mathcal{C}=s_x K$. When all three symmetries are considered, the only possible perturbation that can be applied to $H_0$ is $V^\xi =\xi \Delta \sigma_z \mathbf{s}\cdot\hat{\mathbf{v}}$, where $\Delta$ is the strength of the perturbation, and $\hat{\mathbf{v}}$ is a vector in the $xy$-plane that depends on an arbitrary phase factor introduced when defining the hexapolar basis modes $\psi_3^\pm$. Without loss of generality, we may fix the perturbation to $V^\xi=\xi\Delta\sigma_z s_y$, which is equivalent to setting the relative phase between the basis modes to zero (see supplementary information).  

The perturbation $V^\xi$ commutes with the Dirac Hamiltonian $H_0$ and does not open a gap in the spectrum. This behavior contrasts with graphene, where the only perturbation allowed by both time-reversal and inversion symmetry, the spin-orbit interaction, anti-commutes with the Dirac Hamiltonian and opens a gap (see supplementary information for a more detailed comparison of the electronic and plasmonic honeycomb lattices). For the plasmonic honeycomb lattice, after the $V^\xi$ perturbation is applied, the energy dispersion around the $\mathrm{K}$-point becomes $E_{1,2}=\Delta \pm \hbar v q$, $E_{3,4}=-\Delta\pm\hbar vq$ which is two Dirac cones shifted in energy by $\pm \Delta$ (see \cref{fig:1}d). A dispersionless nodal line is formed at zero energy due to the symmetric shifting of the Dirac cones. Since $V^\xi\propto \sigma_z s_y$ commutes with $H_0$, the eigenstates of $\sigma_z s_y$ are also eigenstates of the full Hamiltonian. The eigenstates of $\sigma_z s_y$ are $|A,\tilde{\psi}_{+3}\rangle$, $|B,\tilde{\psi}_{-3}\rangle$ with eigenvalue $\lambda=+1$ and $|A,\tilde{\psi}_{-3}\rangle$, $|B,\tilde{\psi}_{+3}\rangle$ with eigenvalue $\lambda=-1$, where we have defined the basis states $|\tilde{\psi}_{\pm 3}\rangle = |\psi_3^+\rangle \pm i|\psi_3^-\rangle$. These new basis states have an angular dependence of $|\tilde{\psi}_{\pm 3}\rangle\propto e^{\pm i3\theta}$ and may be referred to as angular momentum basis states. Therefore the Dirac cone shifted by $+\Delta$($-\Delta$) will be formed by states with $\lambda=+1$($-1$). The nodal line at zero energy is formed by a crossing between states in orthogonal subspaces of $\sigma_z s_y$ with opposite values of $\lambda$. The band structure and eigenstates at the K$'$-point are given by the application of $\T$. 

 \begin{figure*}
 	\centering
 	\includegraphics{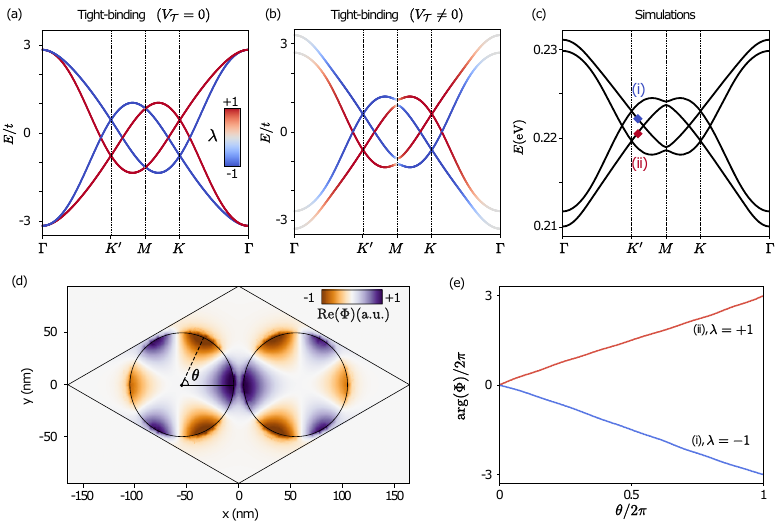}
	\caption{\textbf{Symmetry-enforced band crossings and a $\mathcal{T}$-breaking perturbation}. (a) Tight-binding band structure along high-symmetry lines when all symmetries of the lattice are preserved. The color of the bands represents the value of $\lambda$. (b) Tight-binding band structure with the $\mathcal{T}$-breaking perturbation included. (c) Band structure from full-wave simulations. The real part of the electric potential for the state labeled (i) is shown in (d). The phase as a function of $\theta$ as defined in panel (d) for the states labeled (i) and (ii) are shown in (e).}
 	\label{fig:2}
 \end{figure*}
{\color{NavyBlue}\emph{Tight-binding model and simulation results}}.---A deeper understanding of the plasmon band structure and the physical origin of the perturbation $V^\xi$ can be obtained from a tight-binding model. The plasmonic lattice can be mapped onto a tight-binding model where the hopping coefficients are calculated using an overlap integral between the localized eigenstates of the metallic disk (see supplementary information). As noted previously, the nearest-neighbor hopping coefficients $t_\pm$ between the $\psi_3^\pm$ modes have opposite signs, i.e. $t_+>0$ and $t_-<0$. When $t_+=-t_-$, the tight-binding Hamiltonian including only nearest-neighbor interactions is given as
\begin{equation}
\begin{aligned}
	H_0(\mathbf{k})=t\big[&(1+\cos(\mathbf{k\cdot a}_1)+\cos(\mathbf{k\cdot a}_2))\sigma_x \\ &-(\sin(\mathbf{k\cdot a}_1)+\sin(\mathbf{k\cdot a}_2))\sigma_y\big]s_z
\end{aligned}	
\end{equation}
where $\mathbf{a}_{1,2}$ are the lattice vectors of the honeycomb lattice. 

The hexapole modes can also couple to quadrupole (octupole) modes at lower (higher) energies. Notably, this inter-orbital interaction results in an effective second-neighbor interaction that couples hexapole modes of opposite spin as shown in \cref{fig:1}b,c. The tight-binding form of the second-neighbor interaction is given as
\begin{equation}
	V(\mathbf{k})=t_S\left[\sum_l\sin(\mathbf{k}\cdot\mathbf{a}'_l)\right]\sigma_z s_y
\end{equation}
where $\mathbf{a}_1'=\mathbf{a}_1$, $\mathbf{a}_2'=-\mathbf{a}_2$, and $\mathbf{a}_3=\mathbf{a}_2-\mathbf{a}_1$. This interaction is precisely the symmetry-allowed term $V^\xi$ that was derived for the continuum model. The $\sin(\mathbf{k}\cdot\mathbf{a}'_l)$ coefficient has opposite signs for the K, K$'$-points as required for the $V^\xi$ interaction. The tight-binding band structure including $V(\mathbf{k})$ is shown in \cref{fig:2}a.

The time-reversal symmetry operation of the system satisfies $\mathcal{T}^2=-1$, ensuring a Kramer's degeneracy at the time-reversal invariant momenta $\Gamma$ and $\mathrm{M}$, where states with opposite $\lambda$ become degenerate. The Kramer's degeneracies are evident in the tight-binding band structure shown in \cref{fig:1}a. At the K and K$'$ points, we have shown that the perturbation $V(\mathbf{k})$ results in Dirac point degeneracies between states with identical $\lambda$ and that there are no other perturbations allowed by symmetry. Therefore, along any continuous path connecting K/K$'$ to $\Gamma$/M, the band degeneracies must swap pairs and have a closed loop nodal line degeneracy around the K/K$'$ points. 

The $\mathcal{T}$-symmetry of the system is broken when nearest-neighbor hopping amplitudes $t_\pm$ have opposite signs but different magnitudes. This introduces a perturbation
\begin{equation}
\begin{aligned}
		V_\mathcal{T}(\mathbf{k})=\frac{\delta t}{2}\big[&(1+\cos(\mathbf{k\cdot a}_1)+\cos(\mathbf{k\cdot a}_2))\sigma_x \\&-(\sin(\mathbf{k\cdot a}_1)+\sin(\mathbf{k\cdot a}_2))\sigma_y\big]s_0
\end{aligned}	
\end{equation}
where $s_0$ is the identity operator in the $\psi_3^\pm$ basis and $\delta t = |t_+|-|t_-|$. Adding the perturbation $V_\mathcal{T}$ to the Hamiltonian lifts the Kramer's degeneracy at the time-reversal invariant momenta as shown in \cref{fig:2}b. However, even in the presence of the $\mathcal{T}$-breaking perturbation, the nodal lines around the K, K$'$ points are preserved.

Full-wave numeric simulations are performed using COMSOL Multiphysics to verify the continuum and tight-binding models developed for the honeycomb plasmonic crystal. Metallic nanodisks with a diameter $d=100$nm are centered on honeycomb lattice sites where the lattice constant is $a=190$nm. Each disk is modeled as a surface conductivity following the Drude model $\sigma=iD/\omega$ where $D$ is the Drude weight. For concreteness, we use the Drude model for graphene ($D=e^2E_F/\pi\hbar^2$, $E_F=0.5$eV) as it is the prototypical two-dimensional material for plasmonics\cite{Jablan2009, Low2014b, DeAbajo2014}. Nonetheless, any metallic material that can be modeled by a surface conductivity (e.g. thin gold nanodisks) will exhibit the same physics. A substrate with dielectric constant $\epsilon_s=2.2$ is assumed. The nodal line degeneracies are observed in the simulated plasmonic band structure shown in \cref{fig:2}c. The real part of the electric potential, as shown in \cref{fig:2}d, verifies that the band structure is indeed formed by coupling between hexapolar modes. The eigenstates have an angular phase dependence of $e^{\pm i3\theta}$, which gives them well-defined values of $\lambda$. The bands forming the nodal line crossing have opposite values of $\lambda$ (see \cref{fig:2}e), which protect the degeneracy. Finally, we observe that the Kramer's degeneracies at the time-reversal invariant momenta are lifted in the simulations, consistent with the $\mathcal{T}$-breaking perturbation $V_\mathcal{T}$ derived for the tight-binding model.

 \begin{figure}
 	\centering
 	\includegraphics{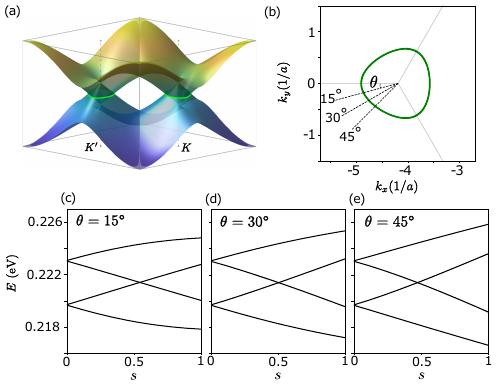}
	\caption{\textbf{Nodal lines enclosing K and K$'$.} (a) Full tight-binding band structure with nodal lines highlighted. (b) Nodal line enclosing the K-point with paths at an angle $\theta$ from the KM high-symmetry line. (c)-(e) Full-wave simulations for $\theta=15^\circ,30^\circ,45^\circ$.}
 	\label{fig:3}
 \end{figure}

The calculations shown in \cref{fig:2} confirmed the existence of protected crossings along the high-symmetry lines of the BZ. However, the nodal lines are expected to form a full loop around the K and K$'$ points. In \cref{fig:3}a, we show the tight-binding band structure for the entire BZ with the nodal lines highlighted. To fully resolve the nodal line, we perform simulations of the band structure along paths that are off the high-symmetry lines of the BZ. The paths all start at the K-point and have an angle $\theta$ to the KM line segment in the BZ (see \cref{fig:3}b). Simulations performed for $\theta=15^\circ, 30^\circ, 45^\circ$ all clearly show the band crossing, confirming that a robust nodal loop indeed encircles the $\mathrm{K}$-point.

 \begin{figure}
 	\centering
 	\includegraphics{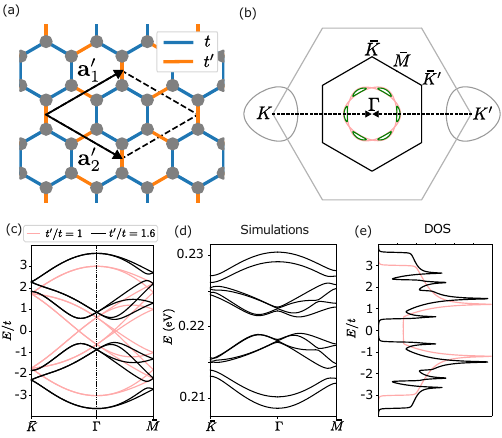}
	\caption{\textbf{Gapping the nodal line with a Kekul\'e distortion}. (a) Schematic representation of a lattice with the Kekul\'e distortion. The dashed lines show the $(\sqrt{3}\times\sqrt{3})$ supercell and the colored bonds represent the modulated nearest neighbor coupling strengths. (b) Folding of the nodal lines at K, K$'$ of the honeycomb lattice onto the $\Gamma$ point. The solid black line represents the BZ of the superlattice while the solid gray line is the BZ of the original honeycomb lattice. Red dashed lines represent the overlapped nodal lines before the bond strengths are modulated ($t'/t=1$) and the green solid lines show the nodal line after a weak modulation is applied. (c) Tight-binding and (d) full-wave simulation band structures for the Kekul\'e distortion. Disks are shifted inwards towards the center of the unit cell by 2nm for the simulations. (e) The density of states is calculated using the tight-binding model.}
 	\label{fig:4}
 \end{figure}
{\color{NavyBlue}\emph{Kekul\'e distortion}}.---Having established how symmetries of the plasmonic honeycomb lattice give rise to nodal lines, we now examine how the nodal lines behave under certain symmetry-breaking perturbations. The case of a $\mathcal{T}$-symmetry breaking perturbation has been discussed above and was shown to lift the Kramer's degeneracies while preserving the nodal lines. A $\mathcal{P}$-symmetry breaking perturbation, which can be realized by making the carrier densities of the nanodisks within a unit cell unequal, gaps the Dirac points but preserves the nodal lines (see supplementary information for details). 

To directly gap the nodal line, we introduce a Kekul\'e distortion, which mixes the K and K$'$ points of the honeycomb lattice. The Kekul\'e distortion requires a $(\sqrt{3}\times\sqrt{3})$ superlattice as shown in \cref{fig:4}a which folds the nodal lines at the K and K$'$ points of the honeycomb lattice onto the $\Gamma$ point. When the nearest neighbor bonds are modulated according to the configuration shown in \cref{fig:4}a, the overlapping nodal lines at the $\Gamma$ point can interact and become gapped. For weak modulations, the nodal lines are only gapped at points of intersection and are broken up into smaller nodal lines (see \cref{fig:4}b). For stronger modulations, the nodal line is entirely removed, leaving a fully gapped plasmonic band structure. The full gap opening is verified using both the tight-binding model (\cref{fig:4}c) and simulations (\cref{fig:4}d). In the simulations, the bond modulation is implemented by moving all disks toward the center of the Kekul\'e unit cell by the same amount. This will have the same effect as the modulation shown in \cref{fig:4}a. 

Interestingly, unlike a gapped nodal point which leaves a two-dimensional parabolic band edge, a gapped nodal loop results in a Mexican hat dispersion. The band edge of the gapped nodal line gives a peak in the density of states that is not observed for gapped nodal points (see \cref{fig:4}e). Since the band edge naturally has zero group velocity in all directions, {the Kekul\'e plasmonic lattice may support slow light with a large optical density of states}. Moreover, the size of the band gap and thus the frequency of the band edge is also tunable by changing the Kekul\'e distortion strength. 

In summary, we have presented a plasmonic honeycomb crystal that supports nodal lines without requiring nonsymmorphic symmetries. The synthetic time-reversal ($\mathcal{T}$), inversion ($\mathcal{P}$), and particle-hole ($\mathcal{C}$) symmetries of the system forbid perturbations that can gap the Dirac node and only allow for an inter-orbital interaction that gives rise to a nodal line around $K,K'$. Such behavior is in stark contrast with electronic honeycomb crystals (e.g. graphene) where a symmetry-allowed spin-orbit interaction gaps the Dirac node. Furthermore, we also show that a Kekul\'e distortion gaps the nodal lines by mixing the $\mathrm{K},\mathrm{K}'$ points of the honeycomb lattice. This is a direct extension of the energy gap opening of Dirac nodal points in graphene induced by Kekul\'e distortions.

Experimentally, localized plasmon modes have been observed in a variety of patterned metallic nanostructures\cite{Yan2012, OBrien2015}. In particular, textured metallic disks have recently been shown to exhibit multipolar resonances and inter-orbital couplings\cite{Li2023}, making it a promising platform to realize the plasmonic honeycomb crystal. While the focus of this work has been on localized plasmonic modes, in general, any platform supporting multipolar electromagnetic modes will be suitable for realizing the artificial honeycomb lattice. For example, localized surface phonon polariton modes in patterned resonators\cite{Chen2014, Caldwell2014} offer the necessary intrinsic multipolar degrees of freedom to realize nodal lines in a phononic honeycomb lattice. 

\begin{acknowledgments}
S.H.P. and T.L. acknowledge support from the Office of Naval Research MURI grant N00014-23-1-2567. E.J.M. was supported by the Department of Energy under Grant No. DE-FG02-84ER45118.
\end{acknowledgments}

 \section*{Methods}
 The full-wave numeric simulations are performed using an eigenfrequency study in COMSOL Multiphysics. The metallic nanodisks are modeled as surface current densities with a Drude conductivity as defined in the main text. A substrate with dielectric constant $\epsilon_s=2.2$ is assumed. For the honeycomb lattice, the unit cell is defined as shown in \cref{fig:1} with periodic boundary conditions along the $\mathbf{a}_{1,2}$ directions. For the Kekul\'e lattice, the unit cell is defined as shown in \cref{fig:4} with periodic boundary conditions along $\mathbf{a}'_{1,2}$. In the out-of-plane direction, the structure is padded by a region of free space with length $10a$ and terminated with a scattering boundary condition.

\end{document}


\title{Supplementary Information: Nodal lines in a honeycomb plasmonic crystal with synthetic spin}

\author{Sang Hyun Park}
\affiliation{Department of Electrical \& Computer Engineering, University of Minnesota, Minneapolis, Minnesota, 55455, USA}
\author{E. J. Mele}
\affiliation{Department of Physics and Astronomy, University of Pennsylvania, Philadelphia, Pennsylvania, 19104, USA}
\author{Tony Low}
\email{tlow@umn.edu}
\affiliation{Department of Electrical \& Computer Engineering, University of Minnesota, Minneapolis, Minnesota, 55455, USA}

\maketitle

\section{Symmetries of the plasmonic and electronic honeycomb lattice}
This section presents a systematic analysis of the symmetries in both plasmonic and electronic honeycomb lattices and which perturbations are thus allowed. The plasmonic and spinful electronic lattice has two spin degrees of freedom and two sublattice degrees of freedom. A general Hamiltonian has 16 degrees of freedom, which can be represented using 5 Dirac matrices ($\Gamma^{a}$), 10 commutators ($\Gamma^{ab}=[\Gamma^a,\Gamma^b]/2i$), and the identity matrix\cite{Kane2005a}
\begin{equation*}
	H(\mathbf k)=\sum _{a=1}^5d_a (\mathbf k)\Gamma ^a+\sum_{a<b=1}^5 d_{ab}(\mathbf k)\Gamma ^{ab}.
\end{equation*}

\subsection{Plasmonic lattice}
For the plasmonic lattice, we may choose the 5 Dirac matrices such that $\T\Gamma^a\T^{-1}=\Gamma^a$ where $\T=i\sigma_z s_yK$. This gives us the following Dirac matrices
\begin{equation*}
	\Gamma^{(1,2,3,4,5)}=\left(\sigma_x s_z, \sigma_z, \sigma_x s_y, \sigma_x s_x, \sigma_y\right).
\end{equation*}
Time-reversal symmetry of the system, $\T H(\mathbf{k}) \T^{-1}=H(-\mathbf{k})$, then requires that $d_a(\mathbf{k})=d_a(-\mathbf{k})$ and $d_{ab}(\mathbf{k})=-d_{ab}(-\mathbf{k})$. Using this notation, the Hamiltonian of a plasmonic lattice with only nearest neighbor interactions may be written as $H_0(\mathbf{k})=d_1(\mathbf{k})\Gamma^1+d_{12}(\mathbf{k})\Gamma^{12}$ where the coefficients are $d_1(\mathbf k)=t\left[1+\cos(\mathbf k \cdot \mathbf a_1)+\cos(\mathbf k \cdot \mathbf a_2)\right],\ d_{12}(\mathbf k)=-t\left[\sin(\mathbf k \cdot \mathbf a_1)+\sin(\mathbf k \cdot \mathbf a_2)\right]$. 

Inversion symmetry is represented as $\mathcal{P}=\sigma_x$ and satisfies $\mathcal{P}H(\mathbf{k})\mathcal{P}^{-1}=H(-\mathbf{k})$. Given that the Hamiltonian satisfies time-reversal symmetry, inversion symmetric terms must satisfy $\mathcal{P}\Gamma^{a}\mathcal{P}^{-1}=\Gamma^a$ or $\mathcal{P}\Gamma^{ab}\mathcal{P}^{-1}=-\Gamma^{ab}$. Particle-hole symmetry is represented as $\mathcal{C}=s_x K$ and satisfies $\mathcal{C} H(\mathbf{k})\mathcal{C}^{-1}=-H(-\mathbf{k})$. Terms that preserve the particle-hole symmetry are given by $\mathcal{C}\Gamma^a \mathcal{C}^{-1}=-\Gamma^a$ and $\mathcal{C}\Gamma^{ab} \mathcal{C}^{-1}=\Gamma^{ab}$. The only terms that preserve all three symmetries ($\T,\mathcal{P},\mathcal{C}$) are $\Gamma^{35}=\sigma_z s_y$ and $\Gamma^{45}=\sigma_z s_x$. Any arbitrary combination of $\sigma_z s_y$ and $\sigma_z s_x$ also preserves the symmetries of the system which leads to the perturbation $V^\xi = \xi \Delta \sigma_z \mathbf{s}\cdot\hat{\mathbf{v}}$ given in the main text. 

\subsection{Electronic lattice}
For the electronic lattice, time-reversal is represented as $\T=i s_y K$ and the 5 Dirac matrices are
\begin{equation*}
	\Gamma^{(1,2,3,4,5)} =\left(\sigma_x, \sigma_z, \sigma_y s_x, \sigma_y s_y, \sigma_y s_z \right).
\end{equation*}
The nearest-neighbor Hamiltonian is $H_0(\mathbf{k})=d_1(\mathbf{k})\Gamma^1+d_{12}(\mathbf{k})\Gamma^{12}$ where the coefficients are $d_1(\mathbf k)=t\left[1+\cos(\mathbf k \cdot \mathbf a_1)+\cos(\mathbf k \cdot \mathbf a_2)\right],\ d_{12}(\mathbf k)=-t\left[\sin(\mathbf k \cdot \mathbf a_1)+\sin(\mathbf k \cdot \mathbf a_2)\right]$. Note that although the Dirac matrix indices are identical to the plasmonic lattice, the actual matrices are not the same. Of the remaining Dirac matrices, the only ones that preserve both time-reversal and inversion symmetry are $\Gamma^{13}=\sigma_z s_x,\Gamma^{14}=\sigma_z s_y,\Gamma^{15}=\sigma_z s_z$. These terms describe the spin-orbit interaction and open a topologically non-trivial gap at the K, K$'$ points in the honeycomb lattice. Choice of $s_{x,y,z}$ is dependent on the choice of Cartesian coordinate basis. 

\section{Hamiltonian of the plasmonic lattice}
This section provides a formal mapping of the plasmonic lattice onto a tight-binding model using a Hamiltonian derived from the electromagnetic equations of motion\cite{Park2024, Jin2017c}. Consider a spatially varying surface conductivity $\sigma(\omega, \mathbf{r})$. The charge density $\rho(\mathbf{r})$, electric potential $\Phi(\mathbf{r})$, and current density $\mathbf{J(r)}$ are described by the following equations
\begin{equation}
	\rho(\mathbf{r})=\frac{1}{i\omega}\nabla\cdot\mathbf{J}(\mathbf{r})=\frac{-1}{i\omega}\nabla\cdot[\sigma(\omega,\mathbf{r})\nabla\Phi(\mathbf{r})],\quad  \Phi=\Phi_{ext} + \int \frac{\rho(\mathbf{r})}{|\mathbf{r-r'}|}d^2r'.
\end{equation}
The plasmonic lattice can be mapped onto a tight-binding model by writing these governing equations into an eigenvalue equation of a Hamiltonian. Assuming a Drude model with spatially varying Fermi energy $\sigma(\omega,\mathbf{r})=i\frac{e^2 E_F}{\pi\hbar^2\omega}=i\frac{e^2}{\pi\hbar}\frac{\omega_F}{\omega}$ allows us to write the governing equations as $\hat{H}\psi =\omega \psi$ where the Hamiltonian $\hat{H}$ and eigenstate $\psi$ are given by
\begin{equation}
\hat{H}=
\begin{pmatrix}
0 & \hat{U}\mathbf{\hat{p}}^T\sqrt{\omega_F(\mathbf{r})} \\ \frac{e^2}{\pi\hbar}\sqrt{\omega_F(\mathbf{r})}\mathbf{\hat{p}} & 0
\end{pmatrix},\ 
\psi=
\begin{pmatrix}
\Phi(\mathbf{r}) \\ \mathbf{J}(\mathbf{r})/\sqrt{\omega_F(\mathbf{r})}
\end{pmatrix}.
\end{equation} 
The Hamiltonian contains the Coulomb operator $\hat{U}[f](\mathbf{r})=\int f(\mathbf{r})/|\mathbf{r-r}'|d^2r'$ and the momentum operator $\hat{\mathbf{p}}=-i\nabla$. 

Written in this form, the Hamiltonian is not Hermitian. However, it may be decomposed into $\hat{H}=\hat{B}^{-1}\hat{A}$ where
\begin{equation}
	\hat{B}^{-1}=
\begin{pmatrix}
\frac{\pi\hbar}{e^2} \hat{V} & 0 \\ 0 & 1
\end{pmatrix}, \quad
\hat{A}=\begin{pmatrix}
0 & \frac{e^2}{\pi\hbar}\hat{\mathbf{p}}^T\sqrt{\omega_F(\mathbf{r})} \\ \frac{e^2}{\pi\hbar}\sqrt{\omega_F(\mathbf{r})}\mathbf{\hat{p}} & 0
\end{pmatrix}.
\end{equation} 
The eigenvalue equation may then be written as a generalized Hermitian eigenproblem $\hat{A}\psi = \omega \hat{B}\psi$. The eigenstates are orthogonal under the generalized inner product $\langle\psi_\nu|\psi_\mu\rangle \equiv \int \psi_\nu^\dagger (\mathbf{r})\hat{B}\psi_\mu(\mathbf{r})d^2r$. A tight-binding hopping amplitude between states $\psi_\mu$ and $\psi_\nu$ may then be calculated by
\begin{equation}
	\langle \psi_\nu|\hat{H}|\psi_\mu\rangle =\int \psi_\nu^\dagger (\mathbf{r})\hat{B}\hat{H}\psi_\mu(\mathbf{r})d^2\mathbf{r}=\int \psi_\nu^\dagger(\mathbf{r})\hat{A}\psi_\mu(\mathbf{r})d^2\mathbf{r}=-i\frac{e^2}{\pi\hbar}\int \left(\mathbf{J}^*_\nu \cdot \nabla\Phi_\mu+\Phi_\nu^*\nabla\cdot\mathbf{J}_\mu \right)d^2r.
\end{equation}

Note that the effective second-neighbor coupling between hexapolar modes is mediated by an inter-orbital interaction with quadrupolar and octupolar states at energies separated from the hexapolar bands. These contributions can be projected onto the hexapolar subspace by using a Schriefer-Wolff transformation. The effective interaction may then be written as
\begin{equation*}
	V_{\mu \mu'} = \frac{1}{2}\sum _\nu W_{\mu\nu} W_{\nu\mu '}\left(\frac{1}{E^0_{\mu}-E^0_{\nu}}+\frac{1}{E^0_{\mu'}-E^0_{\nu}}\right)
\end{equation*}
where $\mu,\mu'$ are indices of the hexapolar basis, $\nu,\nu'$ are indices of the quadrupolar or octupolar basis, and $W_{\mu\nu}$ are matrix elements that can be calculated using the Hamiltonian derived above. 

\section{Effect of relative phase on perturbation $V^\xi$}
The form of perturbation $V^\xi=\xi\Delta \sigma_z\mathbf{s}\cdot\hat{\mathbf{v}}$ has an arbitrary unit vector $\hat{\mathbf{v}}$ that is connected to a phase factor between the basis orbitals $\psi_3^+$ and $\psi_3^-$. To clarify this connection, consider the tight-binding form of the effective second neighbor interaction $V$
\begin{equation*}
	\hat{V}=\sum_{\alpha\neq\beta}\sum_{\langle\langle i,j\rangle\rangle}t_{\alpha\beta,ij}\hat{c}^\dagger_{\alpha i}\hat{c}_{\beta j}
\end{equation*}
where $\hat{c}$ is the annihilation operator for a hexapole mode, $\alpha,\beta=\pm$ are the indices for the degenerate hexapole orbitals, and $i,j$ are the lattice site indices. The hopping amplitude is given by $t_{\alpha\beta,ij}=\langle \psi_{3,i}^\alpha|\hat{V}|\psi_{3,j}^\beta\rangle$. 

First, consider the case where there is no relative phase between the basis states $\psi_3^+$ and $\psi_3^-$. Assuming the electric potential of the basis states to be purely real, the next-nearest neighbor couplings may be written as $\langle \psi_{3,i}^+|\hat{V}|\psi_{3,j}^-\rangle = \langle \psi_{3,j}^-|\hat{V}|\psi_{3,i}^+\rangle=\nu_{ij}t_S$ where $\nu_{ij}$ is $+(-)$ if the path starting at site $i$ and ending on-site $j$ makes a left(right) turn. The Bloch form of the interaction $\hat{V}$ is then
\begin{equation*}
	V(\mathbf{k})=\left[it_S\sum_l\sin(\mathbf{k}\cdot\mathbf{a}'_l)\right](\hat{a}^\dagger_{+,\mathbf{k}}\hat{a}_{-,\mathbf{k}}-\hat{b}^\dagger_{+,\mathbf{k}}\hat{b}_{-,\mathbf{k}})+\mathrm{h.c.}.
\end{equation*}
where $\hat{a}_\pm,\hat{b}_\pm$ are the annihilation operators for hexapolar modes on the A and B sublattices. It is then straightforward to see that the interaction may be written as $V\sim \sigma_z s_y$ where $\boldsymbol{\sigma}$ act on the sublattice degrees of freedom and $\mathbf{s}$ act on the spin degrees of freedom. 

Alternatively, define the basis states such that there is a relative $\pi/2$ phase. The next-nearest neighbor couplings then become $\langle \psi_{3,i}^+|\hat{V}|\psi_{3,j}^-\rangle =i\nu_{ij}t_S$ and the Bloch form of the interaction is written as
\begin{equation*}
	V(\mathbf{k})=\left[t_S\sum_l\sin(\mathbf{k}\cdot\mathbf{a}'_l)\right](\hat{a}^\dagger_{+,\mathbf{k}}\hat{a}_{-,\mathbf{k}}-\hat{b}^\dagger_{+,\mathbf{k}}\hat{b}_{-,\mathbf{k}})+\mathrm{h.c.}.
\end{equation*}
In this case, we find $V\sim \sigma_z s_x$. In general, the perturbation will have the form $V\sim \sigma_z\mathbf{s}\cdot\hat{\mathbf{v}}$ where $\hat{\mathbf{v}}$ is a unit vector in the $xy$-plane. Without loss of generality, we fix the phase factor to zero and use the form $V\sim \sigma_z s_y$ for the interaction. 

\section{Inversion symmetry breaking perturbation}
\begin{figure}
	\centering
	\includegraphics{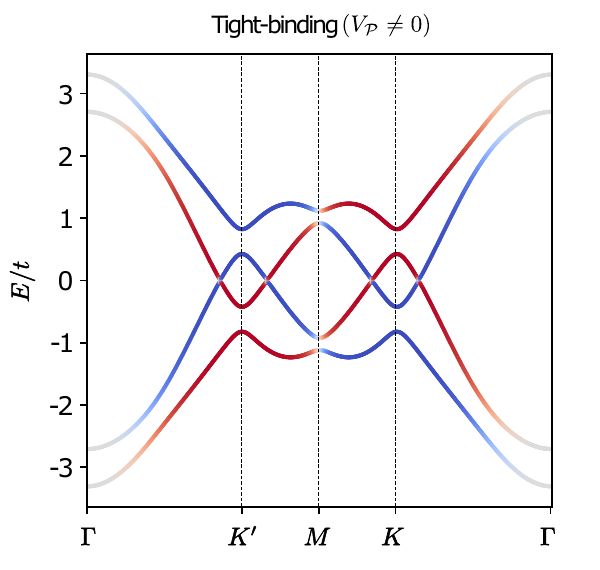}
	\caption{Tight-binding band structure with an inversion breaking perturbation.}
	\label{fig:inversion}
\end{figure}

In the tight-binding model, an inversion symmetry breaking term will have the form of a sublattice staggered potential $V_\mathcal{P}=t_P\sigma_z$ where $t_P$ is the strength of the perturbation. The band structure with both $V_\mathcal{P}$ and $V_\mathcal{T}$ included is shown in \cref{fig:inversion}. The inversion breaking results in a lifting of the Dirac point degeneracy. The synthetic time-reversal breaking lifts the Kramer's degeneracy at the time-reversal invariant momenta as noted in the main text.

\FloatBarrier